# Programmatically Generated Microparticles Using SUEX Dry-Film Epoxy Resist

Jason P. Beech, Jonas O. Tegenfeldt

## Abstract

We present a lithographic method for fabricating free-standing microparticles directly from SUEX dry-film epoxy resist. Unlike conventional SU-8 particle fabrication, which requires patterning on solid substrates followed by sacrificial-layer release, our approach eliminates substrate use entirely and produces particles with near 100% yields. The process supports a wide design space of in-plane geometries, including high-aspect-ratio and highly complex shapes. To enable large-scale particle libraries, we integrate the method with the Nazca Python library, allowing programmatic generation of tens of thousands of parametrically defined particle designs. This combination of substrate-free fabrication and automated design provides a scalable route to custom microparticles for materials science, microfluidics, and soft-matter applications.

## 1. Introduction

While particles have been fabricated in resists such as SU8 (1-3), this requires the use of expensive substrates from which particles need to be released using sacrificial layers together with chemical and/or mechanical treatments, all of which increase material usage, complexity, and cost and often reduce yields. Dry-film epoxy photoresists such as SUEX (4) have predominantly been exploited as structural materials for high–aspect ratio microdevices rather than as free-standing microparticles. Early work by Johnson and co-workers established SUEX dry films as a suitable platform for ultra-thick, high–aspect ratio lithography and LIGA-type structures (5), and demonstrated multilayer patterning and lamination for microfluidic and MEMS applications (6). More recent studies have used SUEX to realize mm-wave ridge gap waveguide components, laminated microfluidic systems with embedded electrodes, and multilayer molds for polymer replication, underscoring its utility for robust, wafer-scale fabrication of micro- and nanostructures (7-9). While freestanding SUEX structures have been reported in the context of laminated microfluidic devices (9, 10), microprobes (11) and flapping wing mechanisms for pico air vehicle applications (12), these examples involve device-scale features rather than discrete microparticles. In all of these cases, however, SUEX serves as a permanent part of the device or as a mold material; to the best of our knowledge, lithographically fabricated, free SUEX microparticles that can be used as discrete colloidal objects have not been reported previously. These particles can be fabricated without substrates, rapidly, with potentially any cross-sectional shape and with yields approaching 100%.

We also present the use of the python library Nazca, to programmatically generate a wide range of particle designs. Using this approach, we can generate series of tens of thousands of particles with almost any 2D shape, provided it can be parametrically described. What is more, we can vary these parameters programmatically if necessary to make each particle unique, either via deterministic or stochastic rules.

This work expands the role of SUEX from a structural device material to a robust medium for high-yield microparticle manufacturing. In combination with automated design generation in Nazca, our approach enables unprecedented freedom in particle geometry and provides a practical foundation for producing large, customized particle libraries for downstream scientific and technological applications.

# 2. Materials and Methods

## 2.1. Generating particle designs

Particle geometries were generated programmatically using Python. The library, Nazca (B. Photonics, - Nazca-Design: Open-source Photonic IC design framework. (2025) - https://nazca-design.org), was used for parametric layout and array generation, while the library gdstk (L. Heitzmann, *gdstk: A Python library for GDSII manipulation*. (2020) - https://heitzmann.github.io/gdstk) was employed for Boolean geometry operations needed to construct irregular particle outlines. These tools enabled systematic control over two-dimensional particle outlines, including size, aspect ratio, and geometric complexity, as well as deterministic and stochastic variation across large particle libraries. All designs were exported in standard GDS II format for photomask generation. Descriptions of the particle geometries and the code that generates them are given in the SI. Figure 2 shows an overview of the methodology used for particle generation and Figure 2A example subsets of the particles fabricated here.

| Shape | Numeric description | Graphic description |
|---|---|---|
| Ellipses | $a = 50\mu m$ <br> $50\mu m \leq b \leq 100\mu m$ <br> $n_{unique} = 100$ | 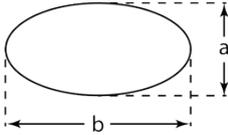 |
| Polygons | $R = 50\mu m$ <br> $n_{sides} = 3, 4, 5, 6$ <br> $n_{unique} = 100$ | 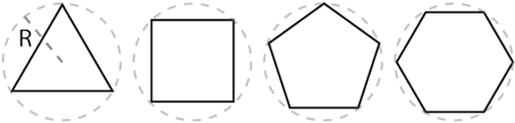 |
| Gaussians | $r(\theta) = R + A exp\left(-\frac{1}{2}\left(\frac{(\theta - \theta_0)}{\sigma}\right)^2\right)$ <br> $R = 50\mu m$ <br> $-49 \leq A \leq 49$ <br> $\sigma = |A| \times 0.2\text{deg}/\mu m$ <br> $n_{unique} = 10\,000$ | 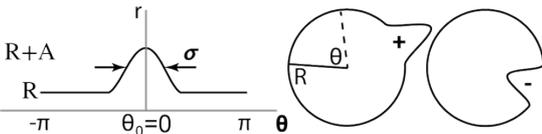 |
| Random Clusters | $shape = \bigcup_i B(\boldsymbol{x}_i, r_i)$ <br> $r_i \sim \mathcal{N}_{\text{trunc}}(\mu, \sigma)$ <br> $\|x_i\text{-}x_j\| \leq r_i\text{-}r_j$ for some $j < i$ <br> $n_{unique} = 10\,000$ | 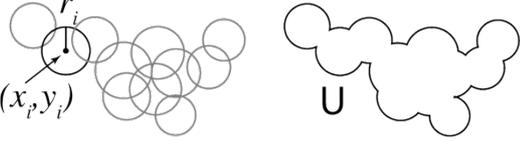 |

*Figure 1 An overview of the particles fabricated in the current work. In the left column the name of the particles, in the middle column a mathematical description with the specific parameters used in the work here, and in the right column schematics to aid in the understanding of particle generation approaches.*

## 2.2.  Fabrication

### 2.2.1. Mask fabrication

Various particle designs were generated in Nazca (see above). 2.5" chrome mask blanks with 540nm of AZ1500 resist were exposed using a maskless lithograpgy system (MLA150, Heidelberg Instruments GmbH, Heidelberg, Germany) at 405nm and a dose of 80mJ/cm$^2$. The mask was developed for 60 s in AZ351B (MicroChemicals GmbH, Ulm, Germany) (1:4 in milliQ water) and dried with nitrogen gas. Chrome was etched for 80 s (Chrome Etch 18, micro resist technology GmbH, Germany ) and remaining resist removed by sonication in acetone and IPA rinse.

### 2.2.2. SUEX Lithography

An overview of fabrication steps is shown in Figure 2. SUEX (SUEX®, DJ Microlaminates, Sudbury, Massachusetts, USA) can be purchased in many sheet sizes. We used sheets cut for 4" wafers and so the film, with protective PET film in place, was cut with scissors into pieces that fit under the area of the mask containing the design and placed on the substrate holder in a lithography system (Karl Suss MJB4 soft UV, Munich, Germany) with the glossy PET film on top. The mask was placed on top, brought into close proximity to the SUEX film and exposure performed at a dose dependent on the resist film thickness (for SUEX K50, 375nm light for 34 s at a lamp power of 30 mW/cm$^2$ gave good results, see Table S1 in the supporting information for more details). The SUEX piece, with PET films still attached were place in a convection oven at 80°C for 10 minutes. After cooling, particles were developed in the following way. The PET film was removed from both sides and the exposed resist placed in a centrifuge tube, covered with mr-DEV 600 (Micro Resist Technology GmbH, Berlin, Germany) and sonicated for 5 minutes. The process of handling SUEX sheets for the removal of the PET films is relatively easy for thicknesses of 50µm and up but for thinner sheets one side of the PET film can be left in place. The developed particles can either be left to sediment to the bottom of the tube or centrifuged for faster processing depending on particle sizes. The supernatant (developer) can then be removed and replaced with IPA and further sonication performed. This process can be repeated until the particles free of unexposed resist and developer and are finally suspended in the desired solvent, at the required purity.

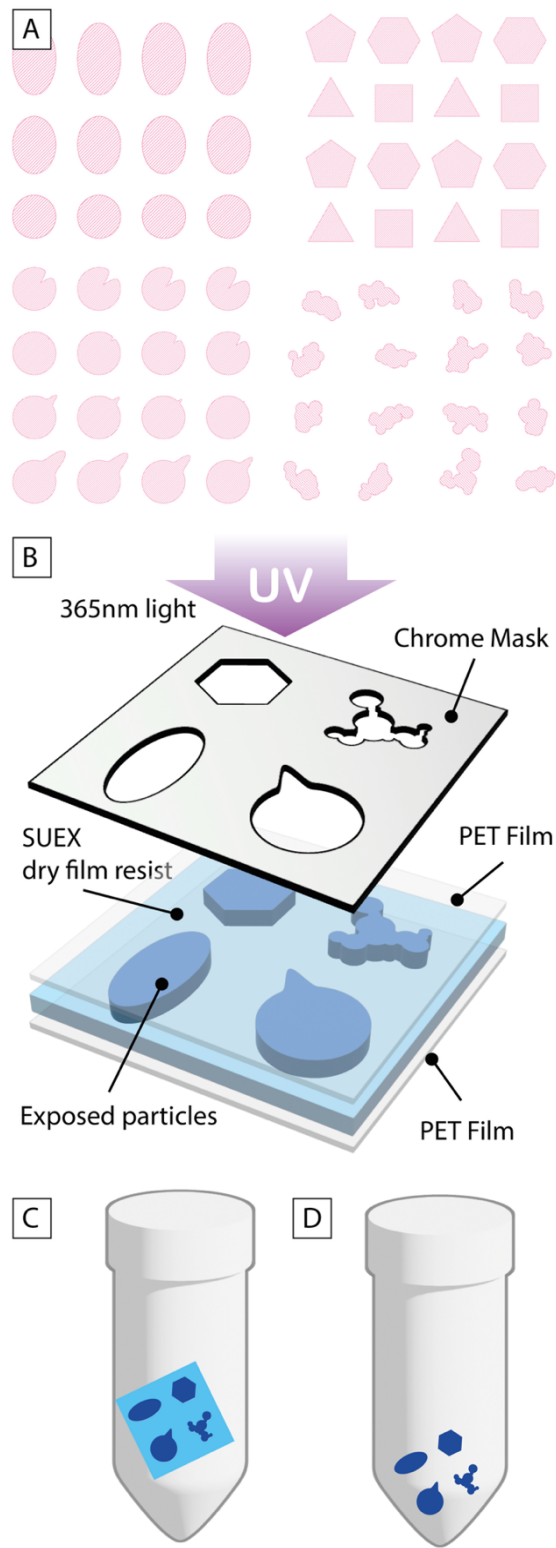

*Figure 2 Design and Fabrication of particles entails designing particlees programmatically with Python. (A) Shows examples of the particles fabricated here. (B) Mask is fabricated and a dry film resist (SUEX) is exposed through the chrome mask. (C) Particles are developed in a tube. (D) After rinsing and liquid exchange particles can be recovered.*

# 3. Results

## 3.1. Various Particle Geometries

Several different types of particles were fabricated. Figure 3 A shows a piece of SUEX K50 film after exposure. The fact that exposed structures are visible would make it possible to do alignment for 2.5D patterning of particles if necessary, which would be possible via multilayer/multiexposure lithography. Figure 3B shows a micrograph of the same exposed SUEX film. Figure 3C shows the exposed SUEX film with the PET sheets removed and placed in a tube ready for development. Static electric charge can make it difficult to handle SUEX sheets and transfer them into the tubes. Using a zerostat gun helps considerably with this. Figure 3D shows approximately 60 000 particles collected in the bottom of the tube after development. Figure 3E-H show the resulting particles imaged in a microscope.

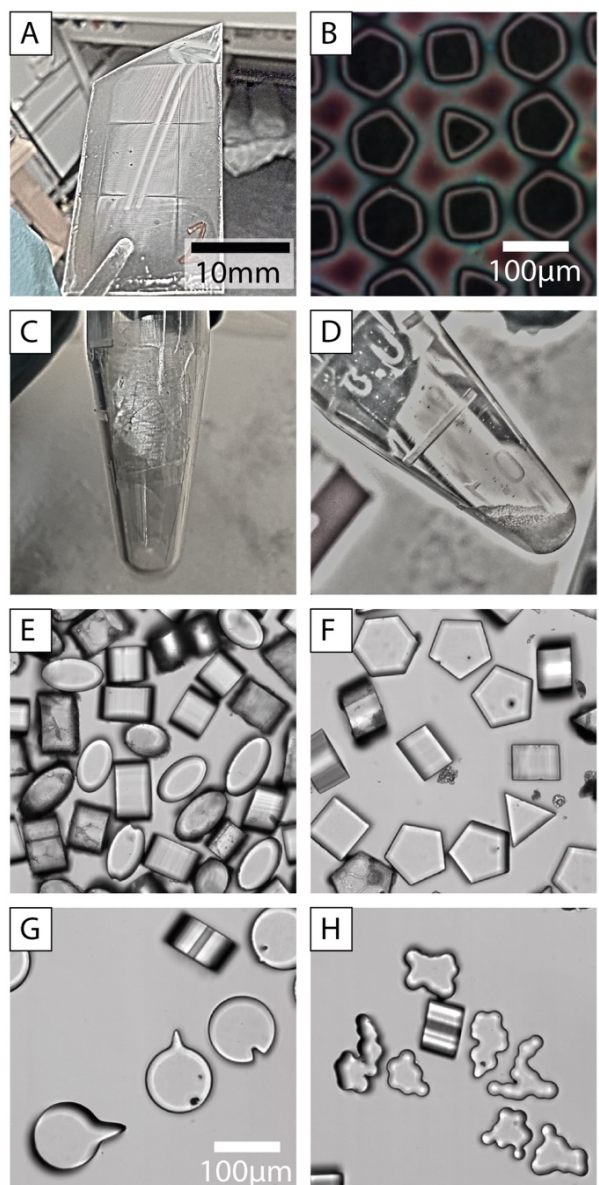

*Figure 3 Particle fabrication steps. (A) SUEX K50 with PET film in place with visible arrays of polygon particles after exposure. (B) Exposed SUEX film imaged through a microscope. (C) Protective PET film removed and resist placed in a tube for development. (D) After development and rinsing, particles collect in the bottom of the tube. (E-H) Examples of particles after development.*

## 3.2. Varying SUEX Film Thickness

SUEX dry-film photoresist is commercially available in a range of laminate thicknesses, enabling straightforward control over particle height with only small modifications to processing conditions. In the following only small changes to the dose were made. We fabricated polygon-shaped microparticles using SUEX K20, K50, K100, and K200 sheets, corresponding to nominal thicknesses of approximately 20, 50, 100, and 200 μm. Each film type was processed using the same lithographic workflow, and in all cases the patterned features released cleanly as fully intact free-standing particles. This range of film thicknesses allowed us to produce particles spanning aspect ratios from thin, plate-like geometries to substantially thicker micro-objects while maintaining consistent shape fidelity and ~100% yield, see Figure 4.

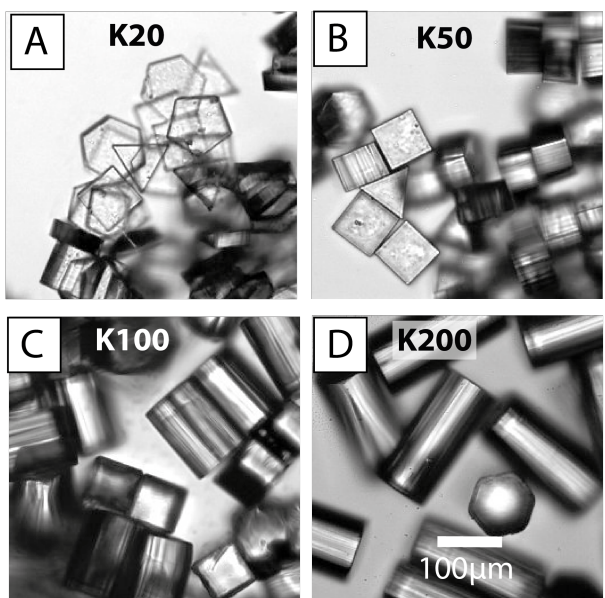

*Figure 4 Microscopy images of polygon particles fabricated with 4 different SUEX film thicknesses. (A) K20, 20µm thick. (B) K50, 50µm thick. (C) K100, 100µm thick. (D) K200, 200µm thick.*

## 4. Conclusion

We have established a robust and highly reproducible workflow for producing free-standing SUEX microparticles spanning a broad range of planar shapes and thicknesses. By eliminating substrate dependence and release steps, our method yields intact particles with 100% recovery across sizes from tens to hundreds of micrometers and for aspect ratios defined solely by the laminated film thickness. Because particles are generated in films that can be freely handled before development, it might also be possible in the future to chemically modify the surfaces of particles and to do this differently on the tops and bottoms of the particles greatly increasing their usefulness. Given access to maskless lithography systems with the correct wavelength for SUEX, it would also be possible to remove the mask fabrication step, which could potentially speed up the fabrication of multiples batches of uniques particles. In summary, the integration of Nazca-based design automation with this SUEX dry film based fabrication approach enables rapid generation of extensive particle libraries, positioning this fabrication strategy as an efficient platform for studies requiring precisely defined micro-objects. We expect to find use for the particles in systematic studies of the effect of shape in self-assembly (13, 14), the behavior of granular matter(15) and for microfluidic sorting(16).

# 5. Declaration of generative AI and AI-assisted technologies in the manuscript preparation process

During the preparation of this work the authors used ChatGPT 5.4 in order for literature research and for assistance with writing code. After using this tool/service, the authors reviewed and edited the content as needed and takes full responsibility for the content of the published article.

# 6. Acknowledgments

We acknowledge financial support by the Swedish Research Council (grant number 2019-04102), and from NanoLund (grant numbers s01-2024, and staff01-2020) and the Crafoord foundation. All device fabrication took place in the cleanroom of Lund Nano Lab at Lund University.

# 7. Credit author statement

**Jason P. Beech:** Conceptualization, Methodology, Software, Validation, Formal Analysis, Investigation, Writing – Original Draft, Writing -Review and Editing, Visualization, **Jonas O. Tegenfeldt:** Conceptualization, Resources, Writing - Review and Editing, Project Administration, Funding Acquisition.

# Supplementary Information for: Programmatically Generated Microparticles Using SUEX Dry-Film Epoxy Resist

Jason P. Beech, Jonas O. Tegenfeldt

## Particle Generation Methods

The following sections provide brief descriptions of the scripts used to generate the particle geometries presented in this work. These summaries outline the purpose and logic of each script; the full source code and version-controlled repository will be made available on GitHub. The example figures show example arrays of 16 particles generated by each of the scripts. For the ectual masks arrays of 100 x 100 particles were generated, but this could easily be increased. ******* JT for the different shapes, I would highlight the unit cell , see example in fig S2 for the polygons. If the unit cell is huge, I would deal with that...

## Ellipses

This script generates an array of ellipse-shaped particle designs with systematically varied aspect ratios using the Nazca layout library. All ellipses share a fixed width, while their heights are scaled to achieve a set of linearly spaced aspect ratios between 1 and a user-defined maximum. Within each unit cell, multiple ellipses are stacked vertically with a constant gap, producing a controlled progression of aspect ratios. These unit cells are then replicated in a $W \times H$ array with uniform spacing in both directions. Each ellipse is rendered as a closed polygon with a prescribed number of points, and the resulting designs are exported as a GDS II file. This layout provides a convenient library of ellipses with controlled geometries for exploring shape-dependent effects in fabrication and experiments.

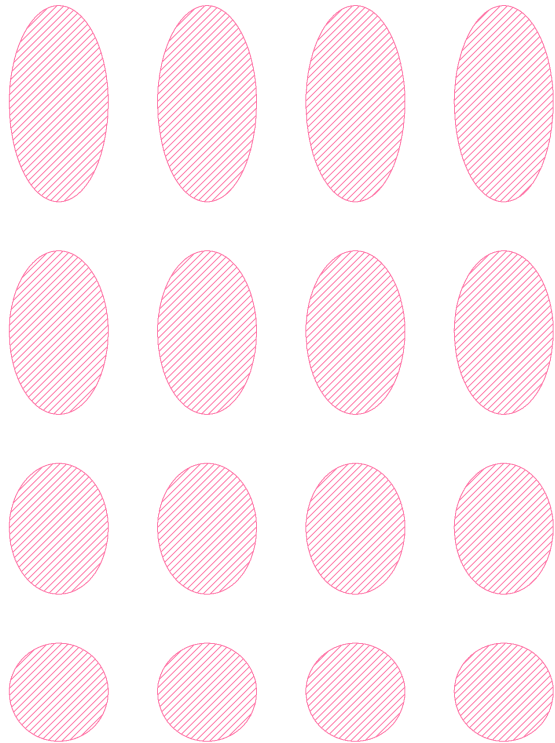

*Figure S1 Small example array of ellipses.*

## Polygons

This script creates a tiled array of "small cells," each containing four regular polygonal particle designs: a triangle, square, pentagon, and hexagon. All shapes share the same circumradius and are positioned on a uniform $2 \times 2$ grid within each cell, with a user-defined gap separating neighboring shapes. The script calculates the center positions for each polygon, applies optional rotations to control orientation, and draws each shape using an $n$-gon generator based on evenly spaced angular samples. Multiple cells are then arranged into an $N \times M$ array using a consistent pitch, ensuring uniform spacing across the layout. The final pattern is exported as a GDS file, providing an efficient method for generating regular-polygon particle sets with consistent size, spacing, and orientation.

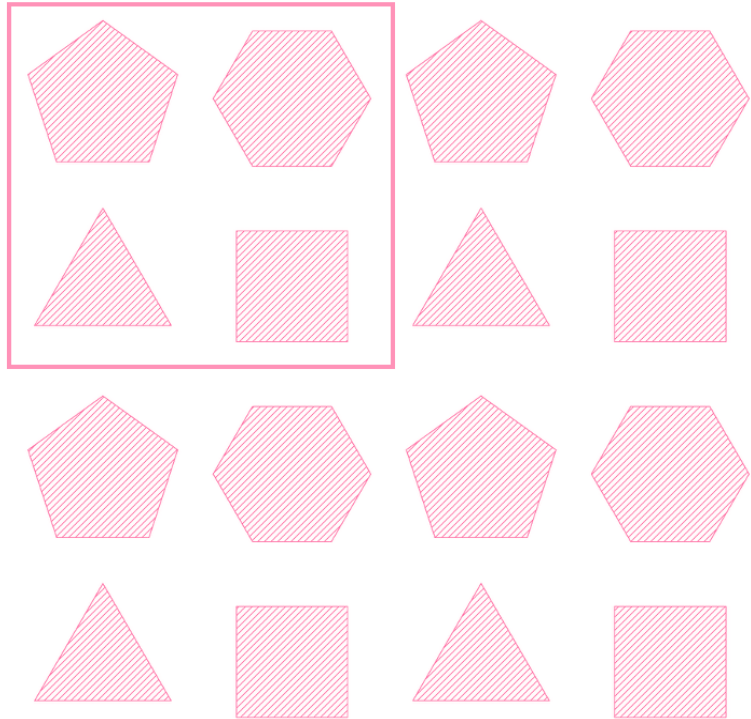

*Figure S2 Small example array of polygons with unit cell highlighted..*

## Gaussian Bump

This script generates an $N \times N$ array of particle designs based on a circular template that is locally deformed by a single angular "Gaussian bump." Each particle is created by sampling a circle at fixed angular resolution and modifying the radius according to a Gaussian function centered at a specified angle. The bump amplitude varies uniformly across the array—largest at the top-left and smallest at the bottom-right—producing a smooth, systematic progression of shapes. The bump width ($\sigma$) is scaled with the amplitude to maintain consistent feature quality across the full range of deformations. Particles are positioned on a regular grid defined by a user-selected pitch, and an optional "snake" mode reverses the indexing of alternate rows without altering the overall amplitude gradient. The final set of shapes is exported as GDS polygons, enabling rapid generation of parametrically defined particle geometries suitable for fabrication.

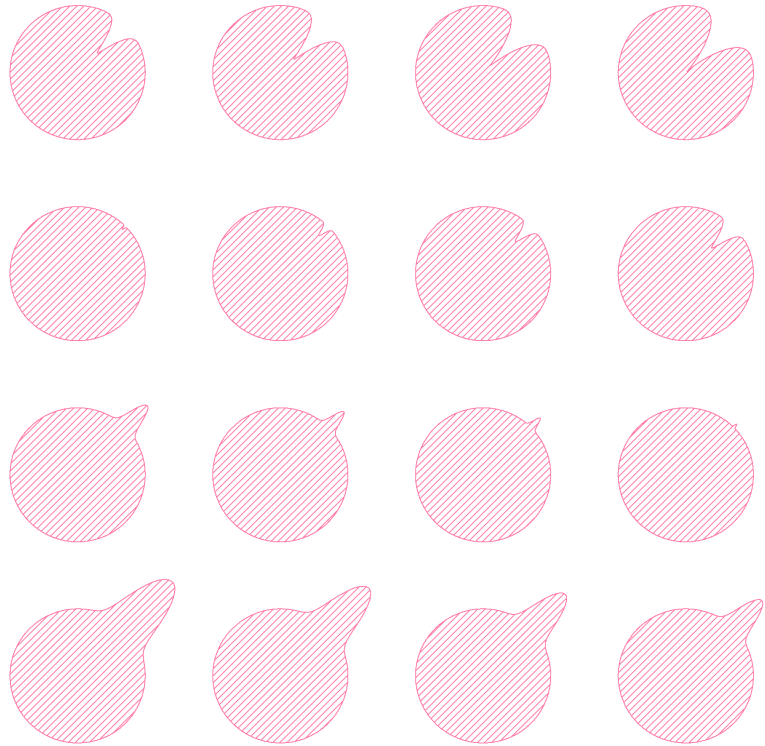

*Figure S3 Small example array of circle with a guassian bump.*

## Random Clusters

This script generates an array of unique, irregular microparticle designs by forming each particle as the geometric union of multiple randomly placed, partially overlapping circles. Circle radii are sampled from a truncated normal distribution, and each new circle is positioned relative to a randomly selected existing "anchor" circle. A placement is accepted only if it increases the overall union area by more than a configurable tolerance, ensuring that each added circle contributes meaningfully to the particle boundary and avoids fully contained circles. Boolean operations are performed using *gdstk* to maintain geometric accuracy, after which the merged outline is converted into polygonal shapes suitable for GDS export.

To assemble an array, the script first generates a sample particle to estimate its bounding-box dimensions and uses this to determine the pitch required to prevent overlap. A full $N_X \times N_Y$ grid is then populated, with each particle independently generated, resulting in hundreds of distinct, topologically irregular shapes. This approach provides a flexible computational pipeline for producing highly diverse microparticle libraries for studies involving disorder, shape variation, and self-assembly.

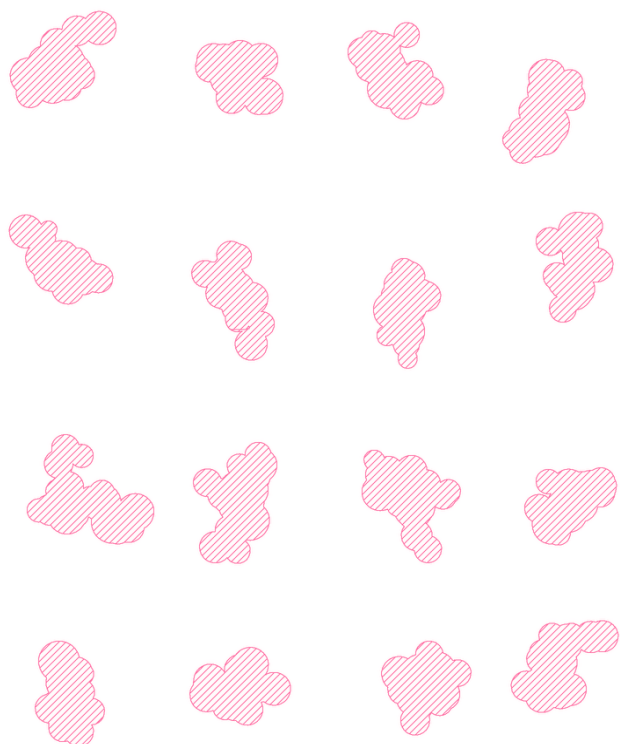

*Figure S4 Small array of random clusters.*

# SUEX Thickness and Exposure Parameters

The lamp power on our system is 30 mW/cm². Exposures were performed in one continuous illumination to give the following doses used in the fabrication of the particles presented here.

*Table S1*

| SUEX sheet name | Nominal thickness [μm] | Exposure dose used [mJ/cm²] |
|---|---|---|
| K20 | 20 | 780 |
| K50 | 50 | 840 |
| K100 | 100 | 1020 |
| K200 | 200 | 1200 |

# Code

Can be found here: doi.org/10.5281/zenodo.19101285